# Local and Global Analysis: Complementary Activities for Increasing the Effectiveness of Requirements Verification and Validation

Lester O. Lobo and James D. Arthur
Department of Computer Science, Virginia Tech
Blacksburg, Virginia 24060, USA
{lester, arthur}@vt.edu

**ABSTRACT.** This paper presents a unique approach to connecting requirements engineering activities into a process framework that can be employed to obtain quality requirements with reduced expenditures of effort and cost. It is well understood that early detection and correction of errors offers the greatest potential for improving requirements quality and avoiding cost overruns in the development of software systems. To realize the maximum benefits of this ideology, we propose a two-phase model that is novel in that it introduces the concept of verification and validation (V&V) early in the requirements life cycle. In the first phase, we perform V&V immediately following the elicitation of requirements for each individually distinct function of the system. Because the first phase focuses on capturing smaller sets of related requirements iteratively, each corresponding V&V activity is better focused for detecting and correcting errors in each requirement set. In the second phase, a complementary verification activity is initiated; the corresponding focus is on the quality of linkages between requirements sets rather than on the requirements within the sets. Consequently, this approach reduces the effort in verification and enhances the focus on the verification task. The second phase also addresses the business concerns collectively, and thereby produces requirements that are not only quality adherent, but are also business compliant. Our approach, unlike other models, has a minimal time delay between the elicitation of requirements and the execution of the V&V activities. Because of this short time gap, the stakeholders have a clearer recollection of the requirements, their context and rationale; this enhances the feedback during the V&V activities. Furthermore, our model includes activities that closely align with the effective requirements engineering processes employed in the software industry. Thus, our approach facilitates a better understanding of the flow of requirements, and provides guidance for the implementation of the requirements engineering process. .

This paper describes a well-defined, two-phase requirements engineering approach that incorporates the principles of early V&V to provide the benefits of reduced costs and enhanced quality requirements.

## 1. INTRODUCTION

The objective of software engineering is to develop and adapt software systems to satisfy user needs, schedule and budget constraints. In pursuit of this goal, a substantial amount of research has been conducted in improving the software development process. However, a majority of the software projects still continue to fail; this is reported in the Standish report which states that only 28% of the real world projects are successful [1]. A primary cause for this state of affairs is attributed to the lack of clear requirements [2]. This finding indicates that the industry still lacks an effective definition of the requirements generation process. The prevailing uncertainty is indicative of the fact that current models inadequately address the requirements phase. Models such as the Requirements Triage [3] and RE Process Framework [4] either address a portion of the requirements process or include abstract, higher level activities that often obscure the implementation aspects of the requirements phase.

The second reason contributing to the poor quality of requirements is that current models accentuate the V&V activities late in the requirements life cycle - immediately prior to the generation of the formal requirements document (SRS). As a result, the V&V activities are burdened with the evaluation of one whole set of requirements at one time which makes it difficult, if not impossible, to focus on the quality of individual requirements. In addition, because the V&V activities are conducted towards the end of the requirements phase, and far removed from the elicitation of requirements, an additional amount of effort must be expended by the stakeholders to revisit the requirements, their context and rationale. Often in this process, there is a loss of information because the requirements become obscured in the minds of the stakeholders because of the large time delay between the requirements elicitation and V&V activities. Our objective is to reduce this time gap, and thereby, introduce and benefit from the clarity of information in the stakeholders' minds. We refer to this closeness of activities as the "nearness in time" factor.

This paper describes a two-phase requirements V&V approach that enhances the quality of recorded requirements, and reduces the time and effort associated with the overall requirements V&V activities. In the first phase, Local Analysis, requirements are incrementally verified and validated based on functional partitions driven by elicitation meetings. This partitioning permits a more focused analysis on individually distinct functions, while maintaining the conventional, quality-oriented V&V activities, but on smaller sets of related requirements. More specifically, V&V activities are executed soon after the elicitation meeting, and before any further requirements are gathered. This strategy facilitates better feedback because the requirements are fresh in the stakeholders' minds.  The second phase, Global Analysis, complements the first and operates on the combined sets of requirements. Building on Phase 1 V&V results, the second phase (consisting of two stages) examines (a) the quality characteristics of completeness, inconsistency and traceability as applied only to linkages among the individual sets of requirements, and (b) business concerns (management, scoping and feasibility) that require a more integrated view of the requirement sets.  The first stage in the Global Analysis phase requires only minimal V&V effort because most of the requirements V&V has been performed in the Local Analysis phase. This enables the requirements engineer to focus on establishing quality relative to the linkages among the requirements sets. Furthermore, the activities addressing business concerns in the second stage of the Global Analysis phase are tightly coupled to provide better focus on analysis and to take advantage of the analyst's clarity of business related information. Taken together, the two-phase approach identifies requirements errors early in the requirements engineering life cycle and supports a more focused subject examination on both requirements quality and business concerns. The consequent effect is a positive impact on the cost and quality of the SRS.

The subsequent sections in this paper explain the two-phase approach for the generation of better quality requirements. In the next section (Section 2), we decompose the Local and Global Analysis phases into their constituent activities and discuss them in detail. Section 3 describes the proposed approach as an integral component of the Requirements Generation Model (RGM) [5]. Finally, Section 4 presents our conclusions and outlines possible future work.

## 2. THE PROCESS

The requirements engineering phase typically consists of elicitation, analysis, specification and verification activities [6]. In this paper, we address these activities with the goal of generating a clear and complete set of requirements in a cost effective manner. Our proposed model focuses on early V&V to alleviate several problems associated with software development [7][8]. The effect is that the cost incurred during product development is minimized through early error detection and correction. Another goal of the model is to represent the linking of requirements engineering activities at a level of decomposition that closely reflects activities in the requirements engineering process practiced in industry today. Thus, our model addresses the inadequacies of many of the earlier approaches that address the requirements engineering phase at more abstract levels, and thus making those models more difficult to understand and implement. Briefly, our approach consists of the following two phases:

- **Local analysis**: an iterative phase concentrating on eliciting, analyzing, documenting, and evaluating small sets of related requirements.
- **Global analysis**: a set of activities that complement the Local Analysis phase and focus on selected verification and related business concerns of the more comprehensive set of requirements.

We will explain the phases in detail in the subsequent sections.

### 2.1 Local Analysis Phase

Local Analysis is the iterative process through which the customer and the requirements engineer discover, review, articulate and evaluate the sets of requirements for the proposed software system. Entering into the Local Analysis phase, we have the Needs Document, which records the customer problems and needs. The Needs Document is generated by the Problem Analysis phase [9] which encompasses understanding the problem, eliciting the stakeholders' needs, and identifying the constraints on the solution. The decomposition of the iterative Local Analysis phase into its constituent activities is illustrated in Figure 1.

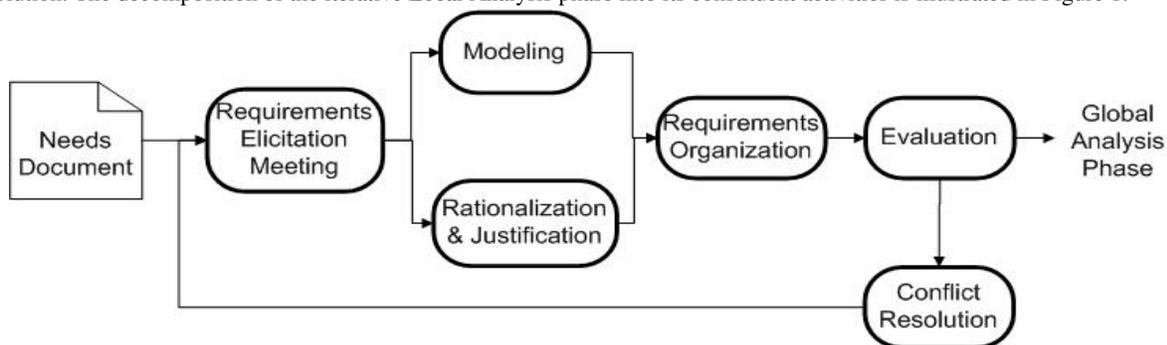

*Figure 1. Local Analysis phase and its activities*

The first activity in the Local Analysis phase is the requirements elicitation meeting whose objective is to correctly identify and capture requirements of the stakeholders. The roles of the requirements engineer and the stakeholder for this activity are complementary - one provides information, the other captures that information. It is the stakeholders' responsibility to convey all necessary system information; it is the requirements engineer's task to elicit and capture the requirements and their context. Several approaches such as Joint Application Design (JAD), Participatory Design (PD) and Facilitated Application Specification Techniques (FAST) have been effectively utilized for this activity. Some commonly used techniques for elicitation are interviews, brainstorming, and focus groups.

To assist in the analysis of the elicited requirements, we introduce the next two activities - Rationalization and Justification, and Modeling - which can be executed either in sequence or in parallel. It is well known that these activities are critical for supporting the analysis of requirements. However, most of the current models such as the Knowledge Level Process Model [10] and the Win-Win model [11] focus on providing a high-level perspective of the requirements process. As a result, lower level activities such as Modeling are often excluded, obscured, or ignored. In turn, this often has a negative impact on requirements generation.

Additionally, the stakeholders often state/identify requirements in an ambiguous manner. Hence, we have incorporated the Rationalization and Justification activity to help identify the reasoning behind the requirements. This activity also analyzes those reasons to determine whether the correct requirements have been identified. If additional requirements are found embedded within the rationale, their relevance and importance are assessed through additional interaction with the stakeholders. This approach continues to foster collective ownership of documented requirements. Identification of the rationale also helps in justifying whether a particular requirement is valid, i.e. the requirement maps to a stakeholder need.

We introduce the Modeling activity to help develop a better understanding of the requirements and to represent them in a clear and comprehensible manner [12]. In addition, the models provide an effective representation for validating requirements with the stakeholders. Furthermore, they are valuable during the Global Analysis and Design phases because those models provide a more precise view of the requirements, their dependencies and their interactions. Hence, the Modeling activity assists in providing a better understanding of the requirements, supports validation, assists in the impact assessment of business concerns, and aids in design of the software system. Various graphical representations, such as Data Flow Diagrams (DFD) and Entity-Relationship Diagrams (ERD) are used in the industry for modeling purposes. These diagrams assist the understanding and formulation of requirements, and are often included in the SRS as supplementary material.

At this point in the Local Analysis phase, the requirements are represented as simple unordered lists. Hence, we have incorporated the Requirements Organization activity to identify important requirements attributes and to structure the requirements for better understanding and analysis. Some of the major requirements attributes are: associated risk factors, effort needed, importance to the user, and value addition to the final product. Requirements Organization also entails the classification of requirements into functional and nonfunctional categories. During the Requirements Organization activity, the functional requirements are further categorized based on the functionalities/features of the system. The non-functional requirements are classified into groups based on performance, security, usability, and so on. The most common techniques used for such classification are affinity analysis and hierarchical decomposition.

Evaluation is the final and most crucial activity for achieving our objectives and in differentiating the Local Analysis approach from other model formulations. This activity includes verification and validation of requirements. Verification at this stage is conducted on the sets of requirements corresponding to individually distinct functions. The objective of verification is to evaluate the requirements for quality characteristics such as non-ambiguity, preciseness, verifiability, and the like. Quality attributes such as completeness, traceability, and consistency that mandate the availability of the complete set of requirements, are only partially evaluated during this particular verification activity. Thus, a large part of requirements verification is achieved during Local Analysis, and thereby reducing the effort of verification in the succeeding Global Analysis phase. Requirements verification is often accomplished through techniques such as inspections, audits, and reviews.

The validation component of the Evaluation activity helps determine whether the requirements satisfy the customer's intent. This activity is focused on smaller sets of related requirements - not the complete set. As a result, the stakeholder is more focused on the activity objective, which in turn yields better feedback. Thus, by performing V&V early in the requirements life cycle we reduce the propagation of requirements errors, enhance the quality of the requirements generated, and thereby reduce the costs incurred during development. Early V&V also takes advantage of the "nearness in time" factor which helps the stakeholder maintain a more focused vision on the recently elicited requirements, and results in better feedback and higher quality requirements. Requirements inconsistencies identified during V&V are addressed in the conflict resolution activity, which is most effective when the interest based bargaining approach is employed [13].

On completion of the Evaluation activity, it is necessary to determine if another iteration of the Local Analysis phase is needed. This decision is based on exit criteria consisting of a checklist of items that pertain to the following: inspecting requirements quality attributes, ensuring that no need/functionality goes unmatched, and finding agreement among the stakeholders that all requirements have been collected. The exit criteria listed here are not comprehensive, but are the necessary items in determining if the transition to the Global Analysis phase can be made.

Several benefits are visible in the proposed approach for Local Analysis and these are outlined below.

- *Incremental and iterative development*: In the Local Analysis phase, the requirements are elicited incrementally based on the functional partitions of the system. This facilitates a better focus on distinct functions because they are analyzed one at a time. In addition, it is also easier to measure progress and validate requirements piecewise.
- *Early V&V*: In a software development lifecycle, a significant portion of the overruns in budget and schedule are a result of having to fix erroneous requirements. Thus, the early detection and correction of requirements errors can significantly reduce problems encountered during software development [8]. We adopt this ideology and perform V&V earlier on smaller sets of related requirements - not on one complete set of requirements towards the end of the requirements phase as prescribed by the current models.
- *Nearness in time*: Because the initial set of V&V activities are executed soon after the elicitation of requirements in the Local Analysis phase, it is easier for the stakeholders to recall the details of individual requirements and their associated contexts. As a result, the stakeholders are more focused and provide better feedback than if V&V were performed at the end of the requirements phase.
- *Better quality requirements*: The early V&V and the "nearness in time" factor enable the generation of better quality requirements early in the requirements phase. This ensures that during the Global Analysis phase we obtain better estimates (cost, schedule, price, risk) which help in making better managerial decisions.
- *Cost effective*: It is well recognized that it is much cheaper to detect and fix errors early in the development life cycle than later [14]. In this model, we do not wait until the creation of the SRS to perform V&V but do so as soon as requirements are elicited. As a result, there is a minimal propagation of errors to the later phases; this positively affects the cost and schedule constraints.
- *Well-defined*: The Local Analysis phase addresses inadequacies found in several current requirements engineering models, i.e., obscured or implied process and activities, or limited scope. The activities of the Local Analysis phase have been selected such that they have clear objectives and provide the seamless evolutionary path for requirements. We have also included two critical analysis activities – Modeling, and Rationalization and Justification – that are often overlooked by current approaches. In addition, we introduce the concept of local conflict resolution to address inconsistencies and incompatibilities in individual sets of requirements.

## 2.2 GLOBAL ANALYSIS PHASE

The objectives of the activities defined in the Global Analysis phase dictate that the elicited requirements be examined as a single comprehensive set rather than as individual subsets. This phase has two overarching objectives: first to complete the requirements verification process and resolve any outstanding conflicts, and second to analyze the requirements from the business perspective. The division of the activities into these two parts is depicted below.

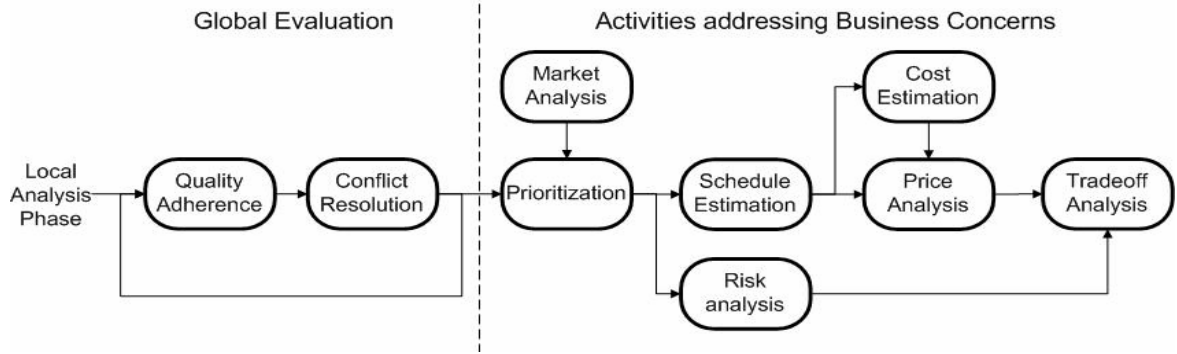

*Figure 2 Global Analysis phase and its activities*

### 2.2.1 Global Evaluation Component

As seen from Figure 2, the first two activities are concerned with a holistic evaluation of the requirements, and are iterative in nature. Because verification during Local Analysis cannot completely address certain quality characteristics (completeness, traceability and inconsistency), we have introduced the Quality Adherence activity during Global Analysis to complement the requirements verification conducted in the preceding Local Analysis phase. Traceability checking, which is included in the Quality Adherence activity, is simplified by the requirements rationale document produced by the prior Rationalization and Justification activity. Coupled with the fact that the Quality Adherence activity concentrates primarily on the linkages between sets of related requirements, with minimal focus on requirements *within* the sets, the effort required to complete the verification process is substantially reduced. Moreover, the concentration on the linkages improves focus and enhances detection of requirements errors. During this verification process, inconsistencies among the requirements are identified through techniques such as inspections, reviews, and audits. Those inconsistencies are resolved through the conflict resolution activity, which should ideally take a participatory rather than the confrontationist approach.

The strengths of the Global Evaluation component of the Global Analysis phase are:

- *Earlier verification*: It is well recognized that early detection and correction of requirements errors positively affects the cost and quality of requirements generated. Unlike many other models, we complete the verification of requirements well before the formal creation of the SRS. That is, the Global Evaluation component verifies the complete set of requirements as soon as they emerge from the Local Analysis phase.
- *Reduced Effort*: Because a major part of verifying quality attributes of requirements is performed during the Local Analysis phase, little effort is needed to complete the verification process, i.e., evaluating the linkages between the sets of requirements.
- *More focused scope*: Because the Global Evaluation component is focused on verifying the links among sets of related requirements, a more thorough examination can be performed.
- *Optional validation*: Although requirements validation is not an objective of the Global Evaluation process, if the customer so desires, validation can be performed to substantiate that the proposed system does reflect its intended purpose.

### 2.2.1 Global Concerns Component

Business concerns are critical in identifying the final set of requirements for the system. Hence, our model focuses exclusively on these concerns in the latter half of the Global Analysis phase. The Business Concerns component helps determine project feasibility and scope, and helps address organizational issues and constraints. The defined activities are an extension to the Requirements Triage model proposed by Alan Davis [3]. In addition, the activities are selected such that they seamlessly synchronize with the rest of our model. During Global Analysis, market information is a critical need because it drives most of the managerial decisions. Hence, we introduce the Market Analysis activity to collect data such as user expectations, market trends, competition, and so forth. The collected data is helpful in determining the price and feasibility of the product. Moreover, to assist in scoping and scheduling the project, we incorporate a Prioritization activity that ranks the requirements based on their importance to the stakeholders and on their value added to the product. Requirements are then categorized into priority groups (low, medium, high priority) or are assigned priorities relative to one another using techniques like the Analytic Hierarchy Process [15].

The remainder of the activities in the Business Concerns component evaluate the requirements from different perspectives to assist in the managerial decisions related to the scope and feasibility of the project. We have grouped these activities together to take advantage of the "nearness in time" effect and to maintain a business concerns focus. The close coupling of activities also promotes better error detection and correction which in turn, positively affects the cost and quality of generated requirements. A brief description of theses activities is listed below:

   a. **Risk Analysis**: focuses on examining the complete set of requirements for risk factors pertaining to product engineering, development environment and program constraints [16]. During this activity, the risk exposure[1] of the requirements is determined. The high risk requirements are documented for discussion with the customer.
   b. **Schedule Estimation**: assists in determining the development time of the components and identifying the critical components of the software system. PERT [17] and CPM[18] are most commonly used for estimating project schedules.
   c. **Cost Estimation**: helps predict the amount of work or effort required in developing the system. The size of the software is a major factor in determining the cost of the project. Size can be represented either in terms of lines of code or function points.
   d. **Price Analysis**: focuses on deciding a fair and reasonable price for the product independent of the cost of individual components and proposed profit. Additionally, this activity also determines which functional capabilities are optional and can be dropped without affecting the value of the product.
   e. **Tradeoff Analysis**: helps evaluate the pros and cons of the system in Operational, Technical, Schedule, Economic, and Legal terms. In addition, tradeoff analysis also includes conflict resolution for requirements that are incompatible with the customer's constraints, e.g., cost, schedule, and profit margins.

Thus, after completion of the second half of the Global Analysis phase, we obtain conflict-free set of requirements that meet customer intent and which have well defined scope. Moreover, the business concerns component supports better management through the generation of timely and appropriately scoped information.

The benefits of the Business Concerns component are:

- *Nearness in time*: The activities evaluating the business concerns are clustered together to help maintain a focused analysis on information related to organizational and management constraints.
- *Early detection of conflicts*: The identification of requirement incompatibilities relative to the business constraints is performed earlier in the requirements phase, that is, it is not postponed until the generation of the formal SRS.
- *Facilitates better management decisions*: Because the quality of the requirements has been established prior to these activities, generated estimates will tend to have less deviation from their actual counterparts. Thus, management can make better informed decisions, and thereby increase the probability of project success.
- *Focused conflict resolution objectives*: Our model establishes a clear distinction among the conflict resolution activities in the Global Analysis and Local Analysis phases. During Local Analysis and Global Evaluation, conflicts in requirements stem from inconsistencies. However, in the latter half of the Global Analysis phase, conflict resolution involves negotiating incompatibilities between requirements and customer constraints. In explicitly recognizing this difference, we are able to apply a more appropriate set of conflict resolution techniques.

## 3. THE RGM

Our modeling approach extends the Requirements Generation Model (RGM) which addresses the complete requirements engineering phase, but from a relatively high-level perspective. The RGM consists of five phases shown in Figure 3.

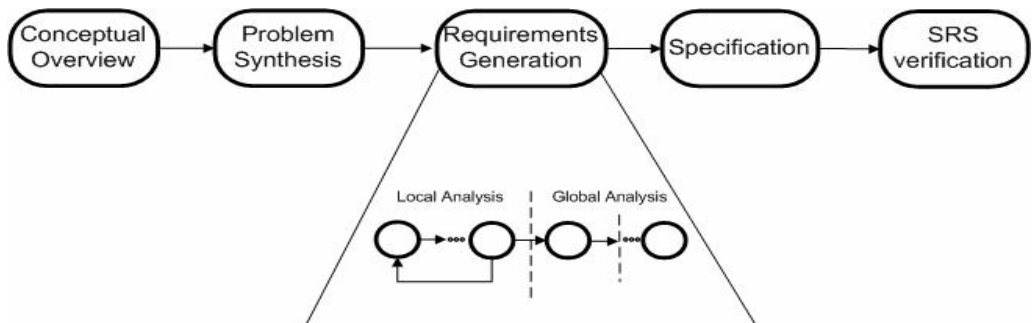

*Figure 3. The local and Global Analysis phases in the context of the RGM*

The first two phases represent the problem domain; they gather information about the existing problems and needs. The remaining phases correspond to the solution domain; they capture the requirements of the solution. The first phase, Conceptual Overview, helps in recognizing the need for the new system both from the business and operational perspectives. The Problem Synthesis phase assists in identifying the customer problems and needs. The Requirements Generation phase helps elicit, analyze and produce a complete set of requirements which adhere to established quality characteristics. In the Specification phase these requirements are then incorporated into a formal SRS. Finally, the SRS is verified and baselined before submitting it to the customer for approval. Our model decomposes the requirements generation phase of the RGM based on the concept of "separation of concerns" [19] which emphasizes the organization and decomposition of the complex process into simpler activities by addressing small sets of concerns. By using the RGM as an extendible basis, we propose that our model integrates seamlessly into the requirements engineering life cycle.

---

[1] Risk exposure is the product of the likelihood that the risk will occur and the magnitude of the consequences of its occurrence.

## 4. CONCLUSION

The model and approach proposed in this paper is novel in that it overcomes several limitations inherent in the previous requirements engineering approaches. It decomposes the requirements phase into activities at a level of abstraction that better illustrates the flow of requirements from one activity to another. Consequently, the resulting decomposition supports a better implementation of the requirements process. Our model and approach also emphasizes several "overlooked" activities (Modeling, Rationalization and Justification) and places them in perspective to the overall requirements generation process. In addition, this model adopts the ideology of early V&V to reduce costs and improve the quality of requirements. Compared to other approaches, our model helps identify and correct errors at the earliest possible time, making the requirements generation process more cost-effective and quality focused. We have designed the model to provide the benefits of the "nearness in time" factor that allows activities to take advantage of the recall clarity. The "nearness in time" factor comes into play both in the local and Global Analysis phases, facilitating better feedback and enhanced product quality. Another factor contributing to the improved quality of requirements is the focus on verifying first the smaller sets of related requirements, and then the linkages among them. Thus, our approach removes the need to verify the complete set of requirements in one attempt. This improves the effectiveness of the verification process and reduces the errors in the final output. Moreover, the verification approach reduces the verification effort when compared to verification performed late in the requirements phase (i.e.) immediately prior to the generation of the SRS. Another benefit of our model is the distinction it makes between conflict resolution relative to requirements inconsistencies and incompatibilities, and to customer constraints. Finally, the model also provides the advantages of an incremental and iterative development approach. Thus, our model, and approach it engenders, promotes the generation of quality requirements in a cost effective manner.

The benefits enunciated above and attributed to our model are substantiated through literature citations and rationalization based on "lessons learned". Nonetheless, we envision a detailed empirical evaluation intended to provide better insights into the implementation aspects and effectiveness of the model. Currently our model focuses on a development process where the customer is an active participant in the requirements generation process, and one that is based on the waterfall approach to software development. We conjecture that its integration into the Object-Oriented and Spiral development approaches can produce similar benefits.